# Deformation Induced Changes in Surface Properties of Polymers Investigated by Scanning Force Microscopy

*S. Hild, A. Rosa, and O. Marti*

*Department of Experimental Physics, University of Ulm, D-89609 Ulm, Germany*

In this study the possibility of combining commercial Scanning Force Microscopes (SFM) with stretching devices for the investigation of microscopic surface changes during stepwise elongation is investigated. Different types of stretching devices have been developed either for Scanning Platform-SFM or for Stand Alone-SFM. Their suitability for the investigation of deformation induced surface changes is demonstrated. A uniaxially oriented polypropylene film is stretched vertically to its extrusion direction. The reorientation of its microfibrillar structure is investigated and correlated to macroscopic structural changes determined by taking a force-elongation curve. Microtome cuts of natural rubber filled with 15 PHR carbon black are stretched. Changes in topography, local stiffness and adhesive force are simultaneously reported by using a new imaging method called Pulsed Force Mode (PFM).

Mechanical properties of polymers, e.g. Young's-modulus or tensile strength, are correlated to their structure. Generally, the structure of a polymer is given by its particular structural arrangement (1, 2). In crystallizable polymers, like isotactic polypropylene, the structural arrangement is given by the arrangement of crystalline and amorphous region. The morphology of crystalline polymers is influenced by the crystallization conditions. During crystallization from the unstressed melt spherolithic structures will grow. The crystallization, which occurs in an external flow field, e. g. during the extrusion process, creates highly oriented structures. Depending on the extrusion conditions lamellar rows, shish-kebab or microfibrillar structures are generated (3, 4, 5). During uniaxial deformation, the differences in mechanical behavior can be observed by taking force-elongation curves (Figure 1). These variations are caused by specific changes in the original microscopic structure during the deformation process. Spherulites are transformed in fibrillar structures (3) or fibrillar structures can reorient (6).

In amorphous especially elastic polymers, like natural rubber, the mechanical properties are mainly influenced by the density of tie-points (7, 9). These tie-points can either be caused by molecular chain entanglements (8) or originated by chemical cross-linking (9), e. g. interlace rubber with sulphur. Another often used technique to modify mechanical properties of rubber is adding carbon black (10, 11). The unique mechanical properties of filled rubber such as increasing mechanical strength if the amount of filler increases are well known on macroscopic scale (Figure 2), but their microscopic origins however still not have been understood in detail.

On macroscopic scale, measuring force-elongation-curves during uniaxial deformation, e. g. stretching, is a conventional technique to determine mechanical properties, like Young's-modulus or tensile strength. To investigate their microscopic origins stretching experiments are combined with other analyzing techniques. Optical microscopy (12), IR-spectroscopy (13) or X-ray diffraction (5, 12) are successfully applied to investigate deformation induced changes in molecular structure, crystallinity or to observe orientation processes. These techniques are mainly suitably to investigate bulk changes. To study changes in surface structure predominately scanning electron microscopy (SEM) (14, 15) is used. For surface analysis by SEM polymers need metallization, because of this it is difficult to investigate more than one deformation state per sample. Scanning force microscopy (SFM) enables to investigate polymer surfaces with respect to their topography and surface properties like local stiffness, adhesion or friction from the Jim-range down to the nm-scale without special pretreatment (16, 17). Therefore the combination of a SFM with a stretching device should provide a new method to investigate surface changes of polymers during uniaxial deformation (18, 19).

Whereas for the conventional analyzing methods like X-ray or SEM different commercial stretching devices are available, up to now no devices are developed for SFM. Few years ago, only SFMs have been available where the sample is scanned (Scanning-Platform-SFM or SP-SFM). These limits size and weight of sample. Because of this, only miniaturized stretching devices can be used. During the last years a new type of SFM, so called Stand-Alone-SFM (SA-SFM), has been developed, where the scanning unit is moved. These microscope can be placed on top of the sample, therefore the limitation in sample dimensions is cancelled. Therefore SA-SFM will be more suitable for the combination with a stretching device than SP-SFM.

In this study we will introduce different types of stretching devices which can either be combined with SP-SFM or SA-SFM. Their suitability for the investigation of deformation induced surface changes of polymers wilt be tested using two samples: A melt extruded Polypropylene film (OPP) is stepwise stretched and deformation induced changes in surface morphology are documented by SFM. Natural rubber filled with carbon black is investigated to document changes in surface topography and surface properties like local stiffness and adhesion during the stretching.

## Instrumentation

In SP-SFM, the size and weight of the stretching device is limited, e. g. in Nanoscope III (Digital Inc., Santa Barbara, CA, USA) the maximum floor space of a stretching device can be 17 mm x 10 mm and the maximum height 10 mm. Because of this, a miniaturized stretching device is constructed, which fits in this SFM. To reduce weight the main parts of the stretching apparatus -two clamping jaws to fix the sample (Figure 3, la and 1b) - are made out of aluminum. One clamping jaw is fixed. The second, mounted on two sliding bars (Figure 3, 2), can be moved by turning a fine pitch thread screw (Figure 3, 3). Pressed the screw towards the fixed part of the apparatus the disengaged part moves. To mount the stretching device into the SFM, the fixed clamping jaw (Figure 3, 1b) is glued onto a magnetic holder. Because of the small size no force sensor and even no stepper motor can be implemented. Using a SA-SFM, this setup enables to measure under water.

The sample, which must have a length of about 20 mm, can be mounted in the device as shown in Figure 3. Two blocks (Figure 3. 4) will fix the sample. A minimum effective sample length (Figure 3, S) of 3.5 mm for stretching is reached, when both clamping jaws are close together. To avoid slipping of the foils during stretching emery paper is glued on this blocks. Because of the SFM dimensions, the maximum elongation which is possible with this setup is 4 mm. By turning the screw elongation steps of 0.1 mm distance can be reached. To determine the stretching ratio the distance between the two clamping jaws is measured by a sliding gauge. The maximum force which can be applied to the sample before slipping takes place is determined to be 5 N. It has been shown, that polymer foils with a thickness major than 20 Jim directly can be imaged. For elastic materials and or thin films, e. g. thin cuts, an additional support in the area of scanning is necessary to avoid vibrations.

Combining a stretching device with a SA-SFM enables to enlarge their dimensions. In Figure 4 a stretching device developed for a commercial SA-SFM {CSEM Inc., Neuchâtel. Switzerland) is shown. Here, stretching is done by two shifting tables driven by computer controlled stepper motors (Figure 4, 1). This balanced assembly is used to secure the symmetric deformation of the sample. Although very slow stretching rates of 0.01 mm/s can be used, no on-line imaging during stretching is possible. This is mainly due to vibrations of the stepper motors. Because of this, for all SFM investigations stepwise stretching and in-situ taking images is necessary. To determine the applied force during the measurement, a force sensor with 5 N range is integrated (Figure 4, 2). This low maximum force is installed for the study of microtome cut sample. To investigate foils or compact samples the force sensor has to be changed. Force values are read out by computer. This enables either to measure force-elongation curves or to control the force during taking SFM images.

To control the force is especially important for the investigation of elastic materials, where relaxation processes can occur when the stretching process is interrupted. In this setup the minimum size of the sample is limited by the dimensions of the cantilever holder (15 mm) of the used SFM. With respect to this size, a maximum elongation of 80 mm is possible. To avoid any sample vibrations during imaging a support (Figure 4, 3) with removable stamp (Figure 4. 4) is mounted below the sample. The height of the support can be adjusted by a micrometer screw. Exchanging the small stamp by heating or cooling plates allows additional control of sample temperature.

For imaging, the SFM is placed on top of three fixed supports (Figure 4, 5). The position of the microscope can be slightly changed. The supports are arranged on the side and in the middle in a way, which the cantilever position is aligned in the center of the sample. The main scan direction is vertical to deformation direction. Although this setup can easily be modified, the large dimensions of the stretching device and the fixed position of the SFM limit the suitability for other commercial SA-SFMs. Therefore, a smaller movable setup has been developed (Figure 5).

In this setup, a SA-SFM can be placed on top of the apparatus by removable supports. If this support is adjusted on x-y-sliding tables, the position of the SFM can be changed during the experiment. This can be necessary if macroscopic changes, e. g. necking, take place outside the region where the cantilever is placed. This compact setup also can be placed on top of an inverse optical microscope. On one hand this allows position control of the cantilever on the other hand this enables the simultaneous analysis of optical properties like birefringence.

## Experimental

To investigate deformation induced morphological changes a semicrystalline material is used. As sample material isotactic polypropylene has been chosen, because its deformation behavior is well studied (3, 4, 6). Two commercial melt drawn polypropylene films (Hoechst, Factory Kalle, Wiesbaden, Germany) are investigated: A uniaxially drawn film (OPP) with a thickness of 50 µm and a biaxially oriented film (BOPP), produced by two-step-stretching process from OPP. with a thickness of 25 µm. The crystallinity, established by Differential Calorimetry, is 67 % for OPP and 71 % BOPP. Both samples show a glass transition at about 12 C.

For the stretching experiments, the OPP film is cut into a rectangle with 20 mm length and 1 mm width. The narrow side is parallel to extrusion direction. After mounting the sample in the miniature stretching device (Figure 3) the effective sample length is 3.5 mm. A force-elongation-curve (Figure 6) for a similar sized sample have measured on a separate stretching device with integrated force sensor. The points of elongation where SFM images have been taken are marked by letters A -D.

To study the surface in the unstrained and strained state, the miniature stretching device (Figure 3) is combined with a commercial SA-SFM (Explorer, Topometrix. Santa Clara. CA. USA). These combination allows in-situ imaging of a stepwise elongated film under water. Imaging under water is necessary for this sample to diminish electrostatic interactions between tip and polymer surface, which can falsify the determined sample height. The SFM measurements have been performed by operating in the static contact mode. As force sensors micro fabricated silicon-nitride ($Si_3N_4$) tips with a spring constant of 0.05 N/m have been used.

Natural rubber (polyisoprene, PI) is used exemplary as elastic material. The investigated PI has an admixture of 1.8 PHR DCP for chemical crosslinking and 15 PHR carbon black N660 (Degussa AG, Hürth, Germany). The average diameter of the carbon black particles is determined to be 67 nm (20). For stretching experiments, these samples are cut using a low-temperature microtome to a thickness of 10 µm. The unstrained dimensions of the thin cuts are 4.0 x 1.0 mm2. Because this is too small for mounting in the stretching device, the slice is glued onto a metallic holder similar to those used for SEM experiments.

The rubber thin-cuts are stepwise stretched to a maximum elongation of 450 %. After each interruption of stretching the sample is relaxed until force is constant, which is about 10 minutes, before taking SFM measurement. The force of the relaxed sample has been recorded. The resulting stress-strain curve is shown in Figure 7. This curve is similar to a curve taken with low deformation rate of about 0.1 mm/sec.

SFM images are taken by a commercial SA-SFM (CSEM inc., Neuchâtel, Switzerland). Because of the softness and stickiness of the rubber surface, imaging in contact mode is difficult. Using Tapping mode (Digital Inc., Santa Barbara. CA. USA) enables to image natural rubber, but no significant difference between filler particles and rubber can be distinguished (21). Due to the high modulation frequency of the cantilever of about 200 kHz rubber appeals hard in Tapping mode. Recently, we have introduced the Pulsed Force Mode (PFM) for the investigation of soft materials without disruptive lateral forces (22). In this mode the cantilever is modulated with frequencies in the range from 500 Hz up to 2 kHz. Because of the lower modulation frequency rubber appears softer than filler particles. Beside this, the PFM enables the additional imaging of surface properties. Topography, local stiffness and adhesive forces are simultaneously mapped. A detailed description of this mode is given in (Rosa, A. Measurm. Sci. Techn., submitted). SFM images of PI thin cuts in unstrained state and strained up to 400 % elongation have been taken using the PFM at a modulation frequency of 1.6 KHz and a modulation amplitude of about 90 nm. $Si_3N_4$-Cantilevers with a spring constant of 0.9 N/m are applied.

# Results and discussion

## Morphological changes of semicrystalline films

In Figure 8 the topography of the unstrained OPP film is shown. In unstrained state the force micrograph is dominated by a fibrillar structure, which is highly oriented parallel to extrusion direction. The lateral extension of these fibrils is larger than the chosen scan range of 2.5 µm. The surface corrugation perpendicular to the fibril orientation is about 25 nm. The average diameter of the fibrils determined by cross-section (Figure 8, A) varies between 20 nm and 50 nm. Imaging the surface with a scan range of 150 µm shows a similar fibrillar structure like Figure 8, but here the diameters of the fibrils are about 200 nm. Because of these observations we assume, that the fibrillar structure detected on 150 µm scan range consists of bundles of smaller, singular fibrils. At the smaller scan range of 2.5 urn these singular fibrils are imaged. The structure of a singular fibril can either be microfibrillar or shish-kebab like (23). Micro fibrils consist of highly oriented polymer chains with a small content of crystalline lamellae. Shish-kebabs are assembled of a nucleus of extended chains with lamellar rows, which grow perpendicular from the extended chains into the melt.

To determine the structure of the singular fibrils illuminated maps are used (Figure 9). In this diagram the fibrils seem to have a weak corrugation vertically to the extrusion direction. Because of these corrugation we assume that their structure is shish-kebab like. Micro fibrils have to have a more or less plain surface. To simplify a singular fibril further will be called "fibril" without differentiation. Figure 9 also shows, that the fibrils are not perfectly aligned one parallel to each other. They are more or less twisted. Although the fibrillar structure can clearly be distinguished, the surface structure appears slightly distorted. We propose to explain this by an amorphous layer on top of the fibrils.

Upon straining the foil perpendicularly to the extrusion direction (ED) a typical force-elongation-curve for semicrystalline polymers with Hookean region, yield point, necking and ductile yielding can be detected (Figure 6). On macroscopic scale during stretching the neck-in of the film and a zone of ductile yielding can be observed (Figure 6). The points of elongation where SFM images have been taken are marked by letters A-D in the force-elongation curve (Figure 6, A-D). Corresponding SFM images are shown in Figure 10, 11, 13 and 14.

Up to 3 % deformation the film shows Hookean elasticity with a linear increase in force. Then the yield point with maximum force of 0.75 N appears in force-elongation curve, either on macroscopic or on microscopic scale changes in the sample shape are visible. Nevertheless on microscopic scale, the interfibrillar areas should be slightly stretched. But these changes in interfibrillar distance are too small to be detected by SFM, because of the tip radius of about 10 nm. Stretching the film closely beyond the yield point, the formation of a neck occurs. Simultaneously the detected force decreases. Now, changes in surface structure can be observed (Figure 10).

The fibrils start to reorient parallel to the applied force. This reorientation process seems to occur stepwise at different surface layers. Whereas on the layers more far away from the surface the parallel alignment still can be seen, on the first layer near to the surface, the parallel alignment of the fibrils is dissolved into a less ordered one. Additionally, at this point of elongation a lot of fibrillar entanglements can be seen. We assume that these fibrillar entanglements will appear at points where former twisting of the fibrils can be observed. Because of this "knots" the failure in between the highly oriented fibrils is prevented during the vertical elongation. Without these strong tie-points, when the fibrillar structure is only connected by amorphous material a film fails if only small load is applied (19).

At further deformation up to 7 % elongation, on macroscopic scale a neck is improved. In the neck region (Figure 6. B) no preferred orientation of the fibrils can be imaged. The whole structure is transformed into a net-like or woven structure (Figure 11).

Based on X-ray and SEM measurements such woven structures have been proposed by several authors {24, 25. 26, 27) to format during the two-step biaxial orientation of polypropylene. In Figure 12 the force micrograph of a biaxially oriented polypropylene film (BOPP) prepared by two step stretching process (Figure 12) is shown.

Again, the layer-like structure of the fibrillar network can be seen. Comparing the illuminated image with the structure of the stretched OPP shown in Figure 11 similar, woven structure can be seen.

After a small increase of elongation up to 10 % the net-like structure is transformed into a deformation state, where the fibrils are again preferentially oriented (Figure 13). Now the preferential orientation is tilted perpendicular to ED. At this strain state also different surface layers can be distinguished.

Stretching the film to elongation rates higher than 10% causes ductile yielding of the sample. The detected force becomes constant. SFM micrographs of 100 % elongated OPP reveal that the fibrillar structures is aligned parallel to the stretching direction (Fig 14), The thickness of the micro fibrils reduced about 20 %. Whereas in Figure 9 the surface seems to be slightly blurred, now the singular fibrils clearly can be distinguished. The amorphous surface layer seems to disappear. An explanation for this effect can be, that the amorphous chains are converted into fibrillar ones.

Based on the SFM images and literature date we propose a schematic model for orientation of the fibrillar structure during stretching perpendicular to extrusion direction, which is shown in figure 15.

When elastic behavior of the film can be observed the interfibrillar areas are slightly stretched (Figure 15, Yield point). During macroscopic necking, the fibrillar structure starts to reorient parallel to the applied force. This reorientation process first happens near to the surface. By way of an intermediate state where no preferred orientation of the fibrils can be observed (Figure 15, Necking) the fibrillar structure is tilting. Further elongation results again in a fibrillar structure where the fibrils are highly oriented parallel to the applied force. This reorientation process is preferred of failure of the film if strong intermolecular tie-points exist. In the investigated film these tie-points seem to be the twisted fibrils, which prevent the film of interfibrillar failure.

## Changes in surface properties

The topography of both unstrained and strained natural rubber filled with 15 PHR carbon black is shown in Figure 16. In unstrained state two regions of height can be distinguished. A low, dark one, which, we assume to be the polymeric matrix. In this dark area higher particles (light) are visible. The maximum height of this features is about 120 nm. Because of their lateral size of about 70 nm, these should be filler particles. After stretching up to 400 % deformation, filler particles are orientated parallel to the stretching direction. Whereas in unstrained state the polymeric matrix appears homogeneous, after stretching a slight line-like structure is visible. These can be due to the orientation of molecular chains parallel to the extrusion direction and reflects the inhomogeneous deformation of the rubber (28).

Changes in surface properties like local stiffness or adhesive force can be used to get more information about the microscopic structure and changes during stretching. First, local stiffness is imaged (Figure 17). In the unstrained state areas of different compliance can be visualized which can be correlated to topography. The main part of the surface appears soft (dark), this correspondents to the area, we assumed to be the polymeric matrix. Because of the chemical structure carbon black particles should have lower compliance than the rubber matrix. The regions with higher topography are harder. This confirmed the assumption that here the filler particles are located.

Whereas in unstrained image the filler clusters show random orientation, they get orientated after deformation to 400 %. The local stiffness of the matrix is reduced. This can be due to the orientation of molecular chains. Like in topography, linear structures parallel to the strain direction appear in the polymeric matrix. But in the local stiffness image these lines are more distinctive than in the topography image. It seemed, that the line-like structures of the matrix connect adjacent filter clusters. Therefore, deformation of the matrix between filler clusters is larger than in the surrounding areas. The rubber is inhomogeneously deformed.

A more careful analysis of the filler panicles shows differences in their compliance. A possible explanation for this observation based on the fact, that the local stiffness signal also contains information of layers below the surface layer (29), Therefore it is difficult in the local stiffness images to distinguish clearly between a surface-near-volume up to a depth of about KM) nm and the very surface itself. It is known that in filled rubber the carbon black is coaled by thin layer of polymer, so called "bound rubber" (30). The differences in the local stiffness are a hint that some of the filler particles should be coated by such a polymeric layer.

A possibility to discriminate between uncoated and coated filler clusters is to investigate the chemical interaction between surface and tip (31, 32, 33) by measuring the adhesive force. Adhesive force is sensitive only for the surface. PFM enables to map adhesive forces simultaneously with topography and local stiffness (Figure 18). In the unstrained sample the polymer matrix shows higher adhesion (lighter) than the filler particles. In the areas where filler particles are presumed two values of adhesive force can be distinguished, which shall be due to uncoated and coated carbon black. Uncoated filler appears black, which means the adhesive force is low. Coated filler shows higher (gray) adhesion, but the adhesive force is lower than in the surrounding matrix. To explain the difference in adhesive force between the polymeric matrix and the bound rubber layer a simple model is used: We propose that the adhesive force is correlated to the formation of adhesive contact between tip and polymer surface. In the unstretched natural rubber the polymer chains are non-ordered like in a spaghetti ball and a lot of free chain ends are available to form an adhesive contacts to the tip. In bound rubber reactive sides of the chains are bound to the carbon black, therefore the number of free chain ends and the probability of forming adhesive contacts, is reduced. The determined adhesive force decreases.

After stretching, the adhesion of the filler clusters is unchanged. Comparing the SFM adhesion micrographs for the unstrained and strained sample, a decrease in adhesive force of the rubber matrix can be observed in such a way, that it is difficult to discriminate between polymeric matrix and coated filler. For a more detailed investigation of the adhesion forces histogram technique is used. Here the amount of pixels belongs to the same intensity, which means the same adhesive force, is determined. In Figure 19 histograms of adhesion images taken at different stain ratios are shown. Two peaks with different intensity are visible: The small one at an adhesive force of about 10 nN does not change the position with increasing elongation. The higher one shifted with increasing deformation towards from a maximum value of 80 nN to lower adhesion 40 nN.

Based on the SFM images the small peak at low adhesive force can be allocated to the carbon black particles. The peak with higher intensity belongs to the rubber. The reduced difference in adhesive force between coated filler and stretched polymeric matrix, confirmed the model of structure dependent adhesive force: During stretching the polymer chains will be oriented. For stretched chains the possibility of adhesive bond formation is reduced. This results in a decrease of measured adhesive force, this model neglected the influence of contact area or contact time. Therefore quantitative analysis of adhesive values cannot be given here.

## Conclusions

This study shows possibilities for the investigation of microscopic surface changes during stepwise elongation by SFM. For isotactic polypropylene, the reorientation of micro fibrils can be visualized and correlated to macroscopic changes detected from the force-elongation-curves. A model for the microscopic surface changes is given. From SFM images it seems that the reorientation processes first starts at the surface. Bulk reorientation appears at higher elongations. For a detailed analysis the experiment has to be carried out again with a stretching device where definite small steps of elongation can be done and the force can be controlled during the whole investigation. This will be possible with the stretching device shown in Figure 5.

Using a new mode, called PFM, enable to investigated deformation induced changes in topography, local stiffness and adhesion of filled rubber simultaneously. The in-homogeneous deformation of the polymeric matrix can be shown. Besides this, filler particles consisting of carbon black are determined. Comparing local stiffness and adhesive force images allows to distinguish the very surface of coated and uncoated fillers. Histograms of adhesive force enables the direct observation of the influence of deformation in the adhesive force. Depending on this a simple model is proposed, which describes the changes in adhesive force by microscopic changes in rubber structure. For a more quantitative description of these phenomenon it is necessary to take into account changes in local stiffness, which results in changes of contact area, and the influence of contact time.

## Acknowledgements

This work was supported by the Sonderforschungsbereich 239. We gratefully acknowledges the Friedrich Ebert Foundation, Bonn, Germany and the DFG for their fellowships. We are thankful to E. Weilandt, B. Zink, R. Brunner, B. Heise, M. Pietralla and H. G. Kilian for fruitful discussions. Sample materials were prepared by M. Kienzle. Control electronics were implemented by G. Volswinkler.

Appeared in Ratner, B. D. & Tsukruk, V. V. (Eds.) "Scanning Probe Microscopy of Polymers" as Chapter 6, American Chemical Society and Oxford University Press, 1998, 694, 110-128## Figure Captions

Figure 1: Schematic diagram of force-elongation curves for different morphologies of polypropylene.

Figure 2: Schematic diagram of force-elongation curves of natural rubber filled with increasing content of carbon black

Figure 3: Schematic diagram of a miniature stretching device for conventional scanning force microscopes.

Figure 4: Stretching device for Stand Alone Scanning Force Microscopes. Deformation parameters are controlled by computer. Integrated force sensor allows direct monitoring of force.

Figure 5: Stretching device suitable for Stand Alone Scanning Force Microscopes. The combination with an inverse microscope allows the position control of sample and cantilever.

Figure 6: Force-elongation-curve taken during vertical stretching of OPP. Macroscopic changes of the foil are shown. Points A-D mark the elongation ratios. where SFM images are taken.

Figure 7: Slress-strain-curve of rubber sample filled with 15 PHR carbon black cut by low-temperature microtome.

Figure 8: The topography image of the unstrained uniaxially oriented polypropylene (OPP) has a maximum height of 25 nm.

Figure 9: For a better visualization of the fibrillar structure of the unstrained uniaxially oriented polypropylene (OPP) illuminated maps can be used.

Figure 10: Illuminated SFM images taken after stretching beyond yieldpoint. Arrow indicates the stretching direction. Reorientation of fibrillar structure starts.

Figure 11: Stretching in direction marked by arrow until macroscopic necking occurs. Illuminated SFM micrograph shows no preferred orientation.

Figure 12: SFM image of biaxially oriented polypropylene (BOPP). The topography shows maximum height of 40 nm (a). For better comparison to OPP images the illuminated map is shown (b).

Figure 13: Stretching near to ductile yielding (10 % in direction of the arrow), reorientation of fibrillar structure parallel to external stress becomes visible in illuminated SFM images.

Figure 14: Illuminated SFM images taken at 100 % elongation when ductile yielding takes place. The fibrillar structure is reoriented parallel to stretching direction, which is indicated by arrow.

Figure 15: Schematic diagram of fibrillar reorientation. Interfibrillar tie molecules prevent failure between adjacent fibrils.

Figure 16: Topography of rubber filled with 15 PHR carbon black. Filler particle appears higher. Stretching up to 400 % causes an orientation parallel to stretching direction (marked by arrow).

Figure 17: SFM micrograph of local stiffness of rubber filled with 15 PHR carbon black. Filler particles are suffer (lighter) than the polymer matrix (dark). After stretching up to 400 % (arrow indicates stretching direction) the local stiffness increases and the line like structure becomes more visible.

p. 10

Figure 18: SFM adhesion micrograph of rubber filled with 15 PHR carbon black. In unstretched state, the polymeric matrix shows high (light), the filler particle low adhesion (dark). After stretching up to 400 %, the adhesion of matrix decreases.

Figure 19: Adhesion distribution of rubber filled with 15 PHR determined at different strain states shows two peaks: The filler particles have an adhesive force at about 10 nN, adhesive force of rubber decrease with respect to deformation ratio.

Figure 1

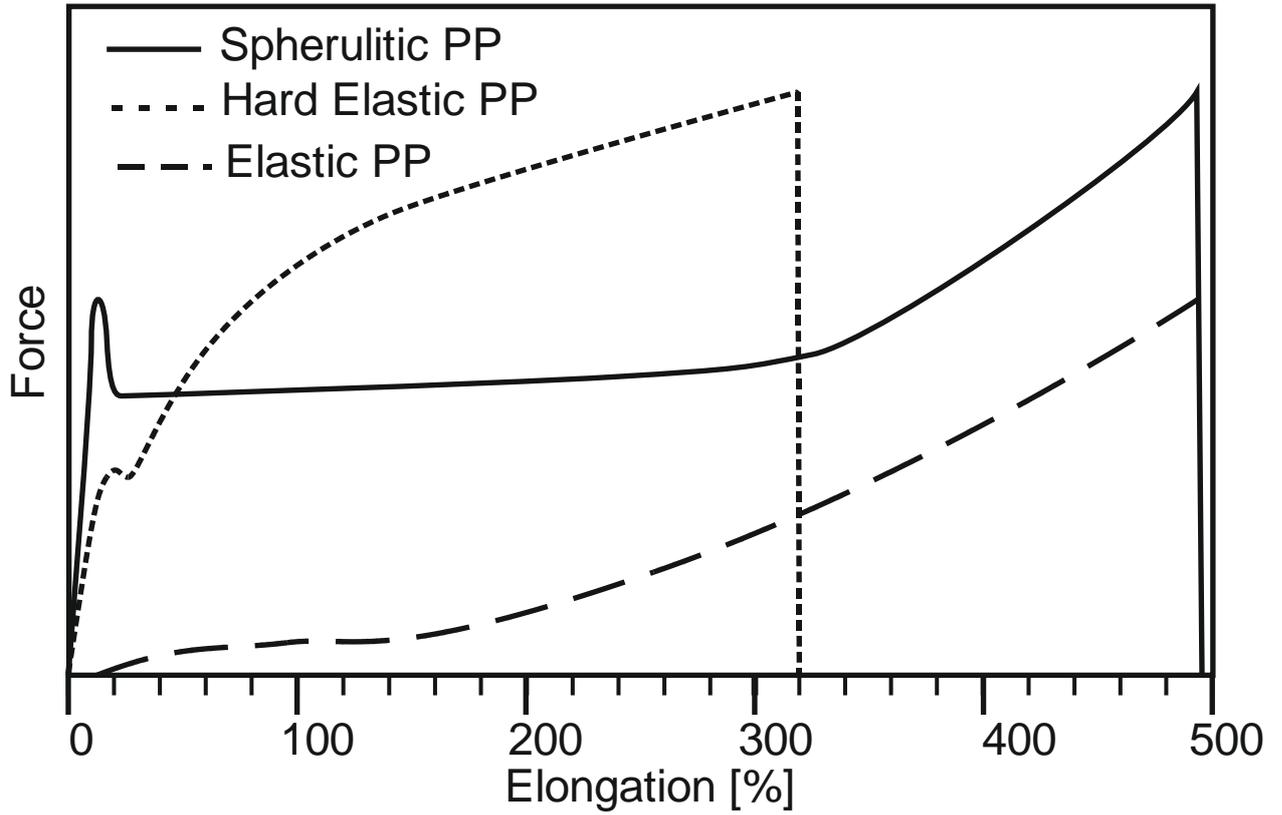

Figure 2

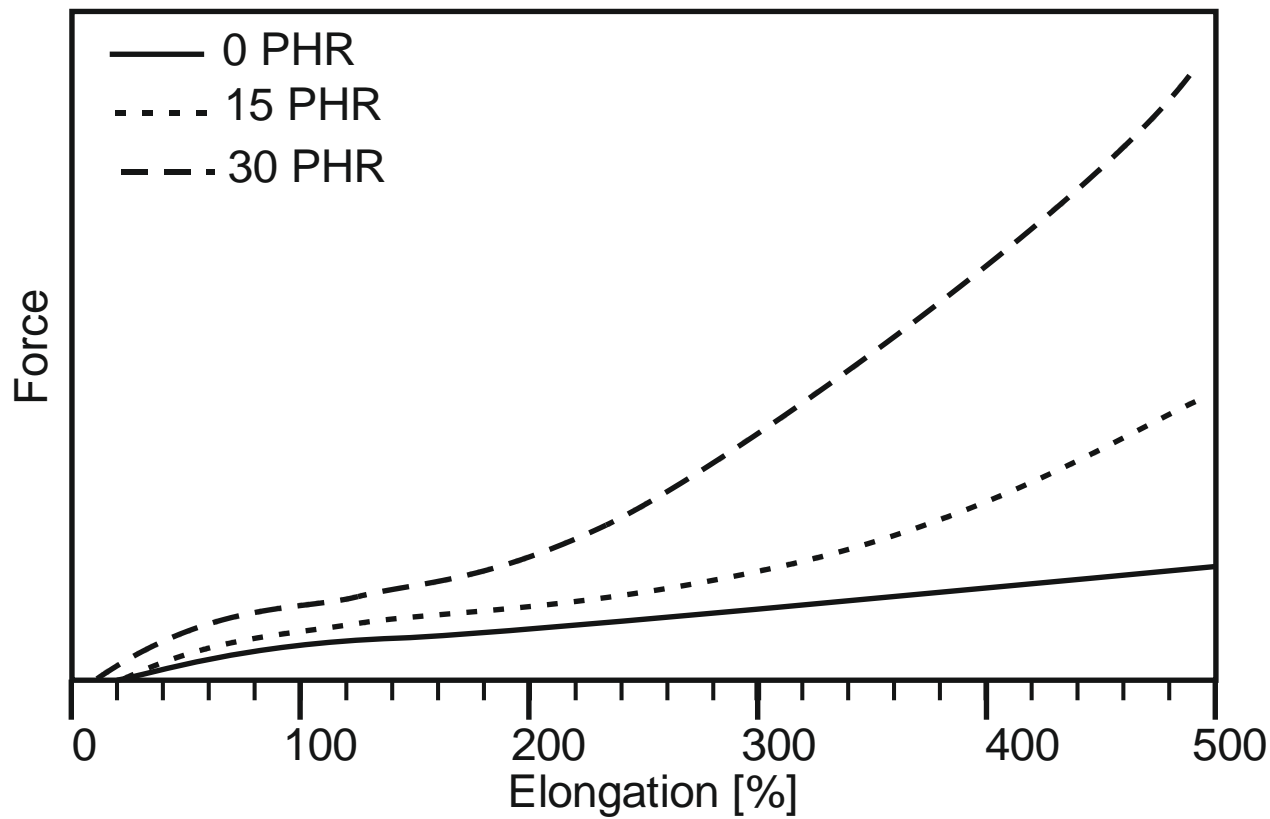

## Figure 3

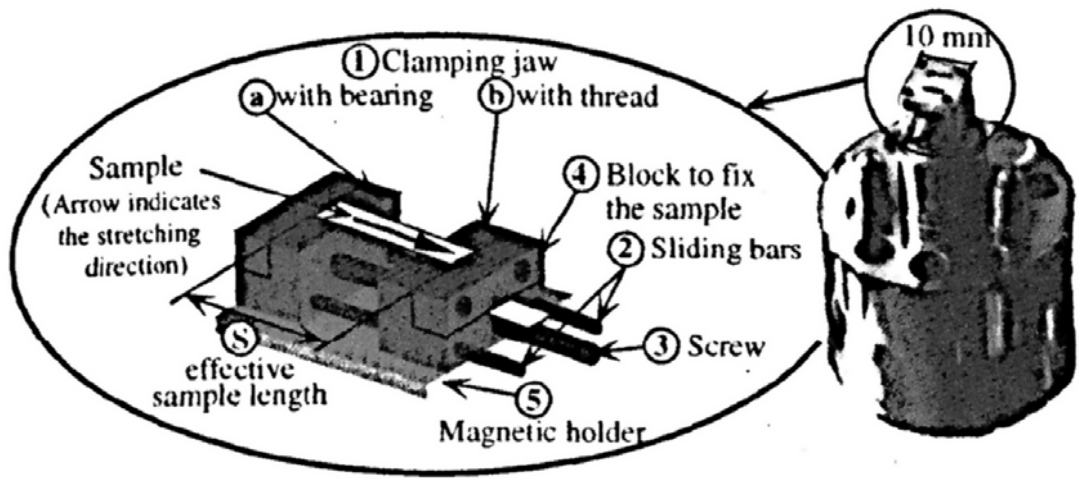

Figure 4

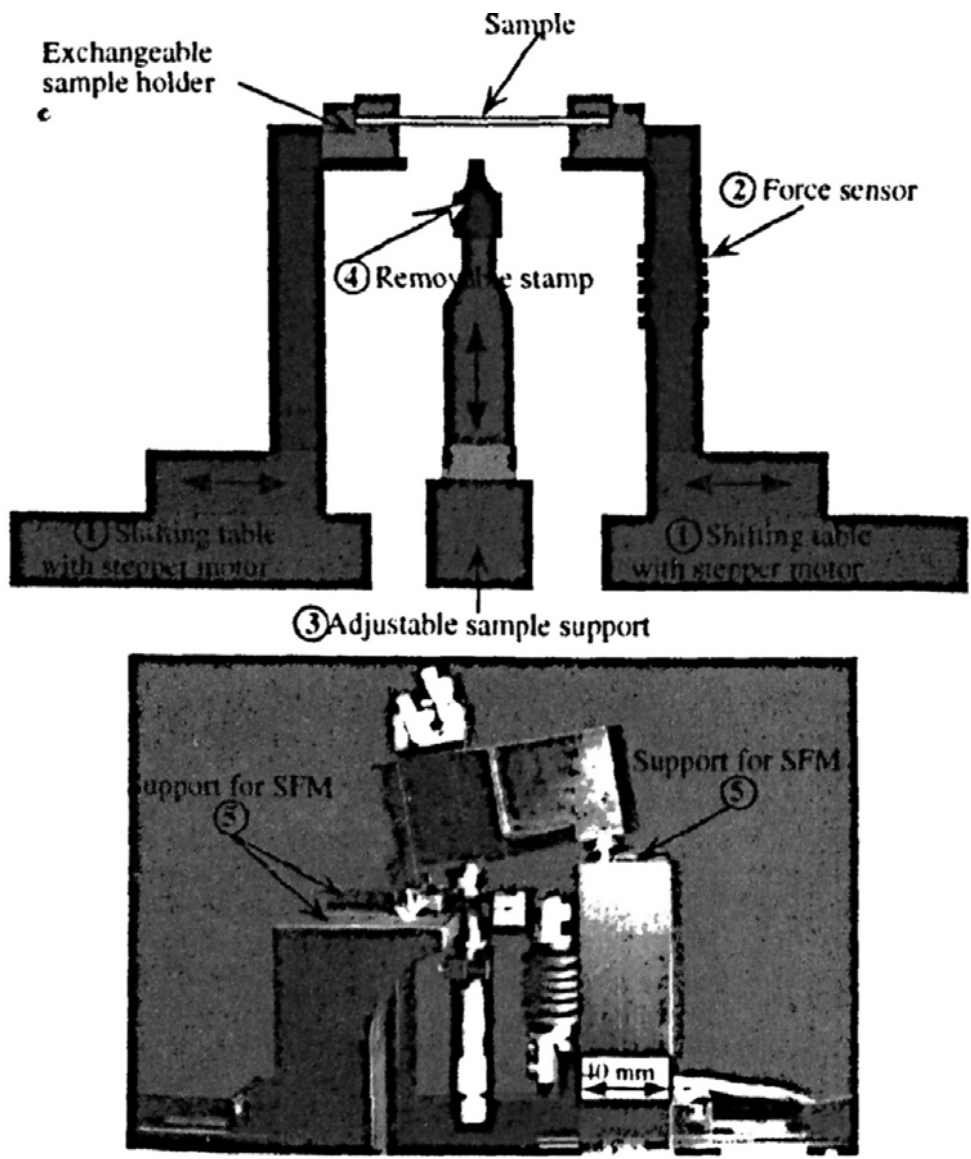

## Figure 5

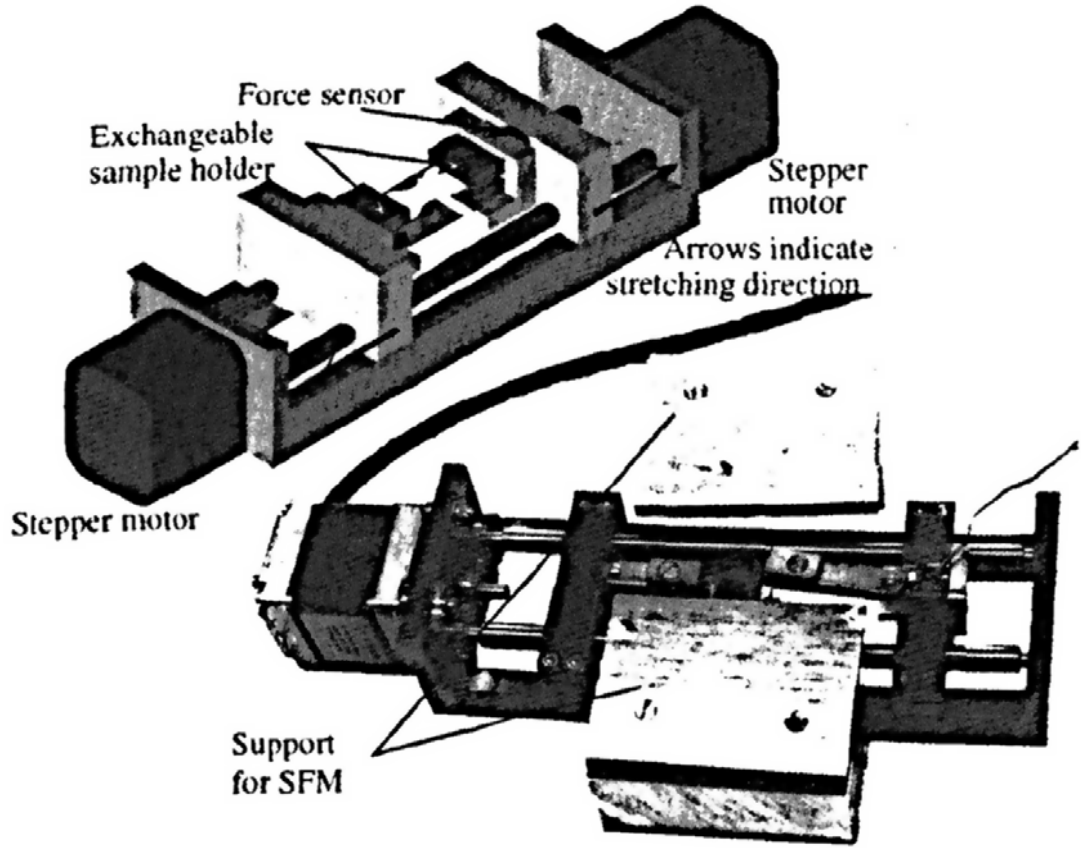

## Figure 6

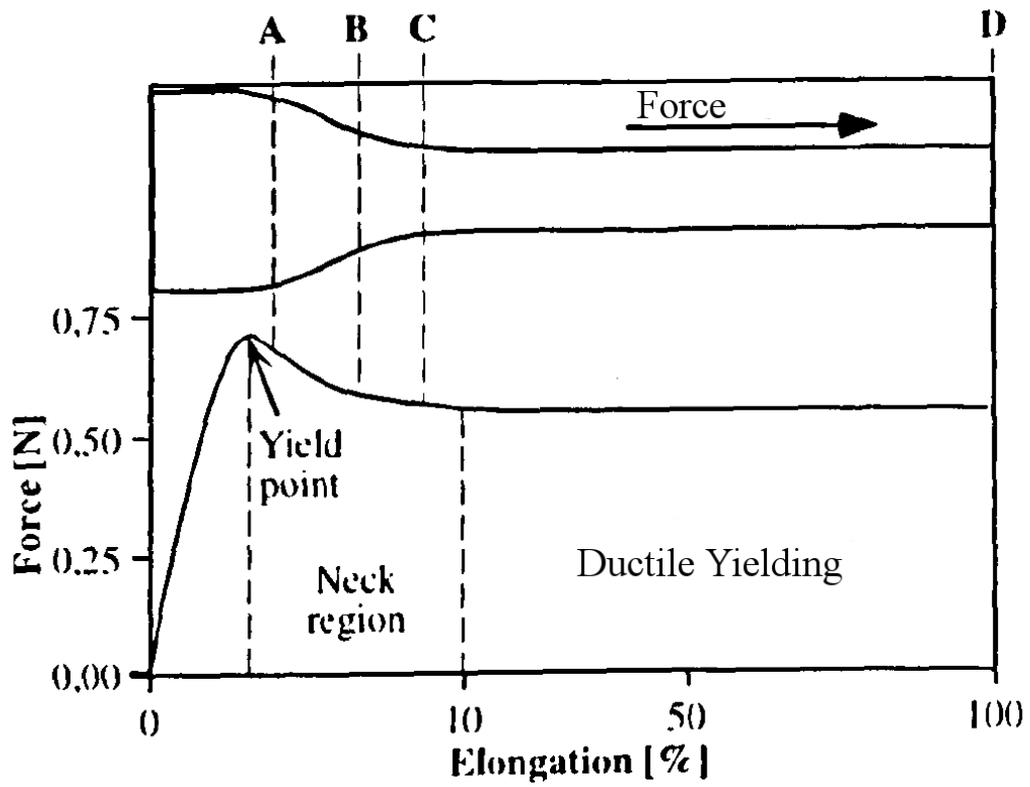

Figure 7

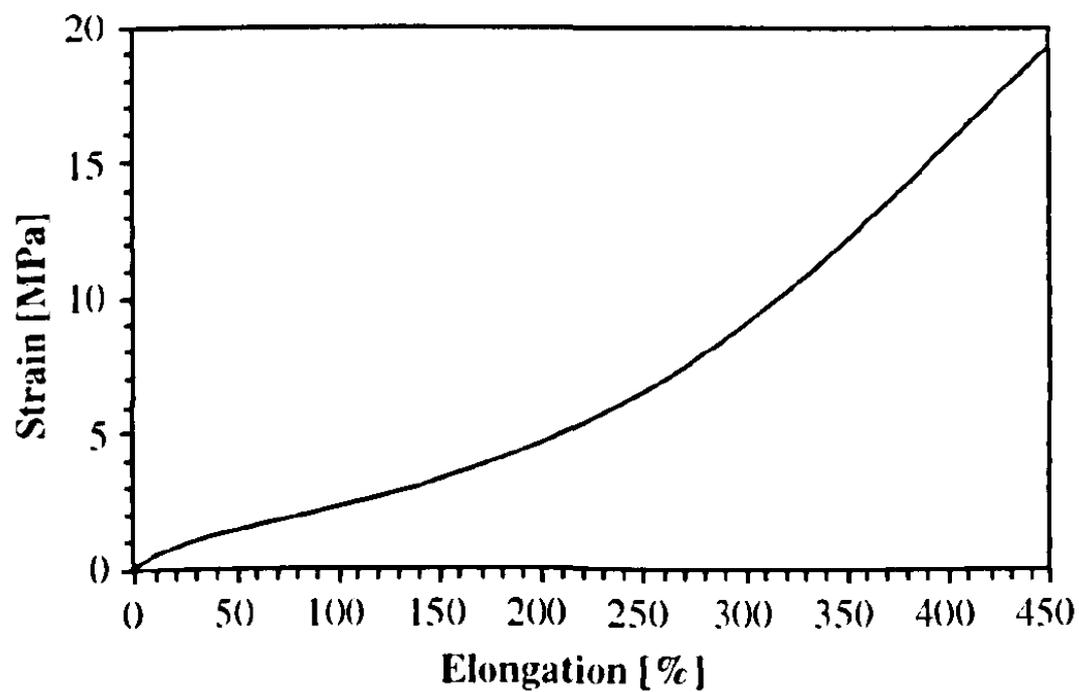

# Figure 8

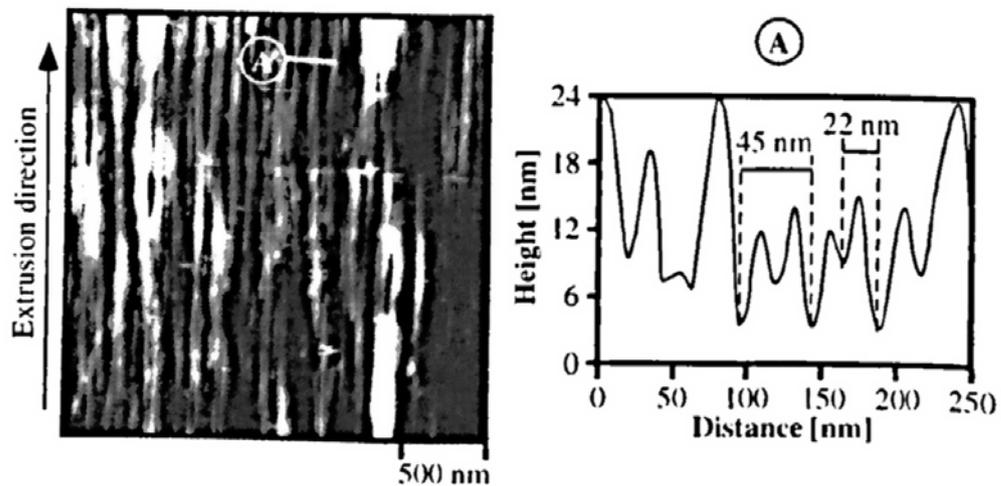

Figure 9

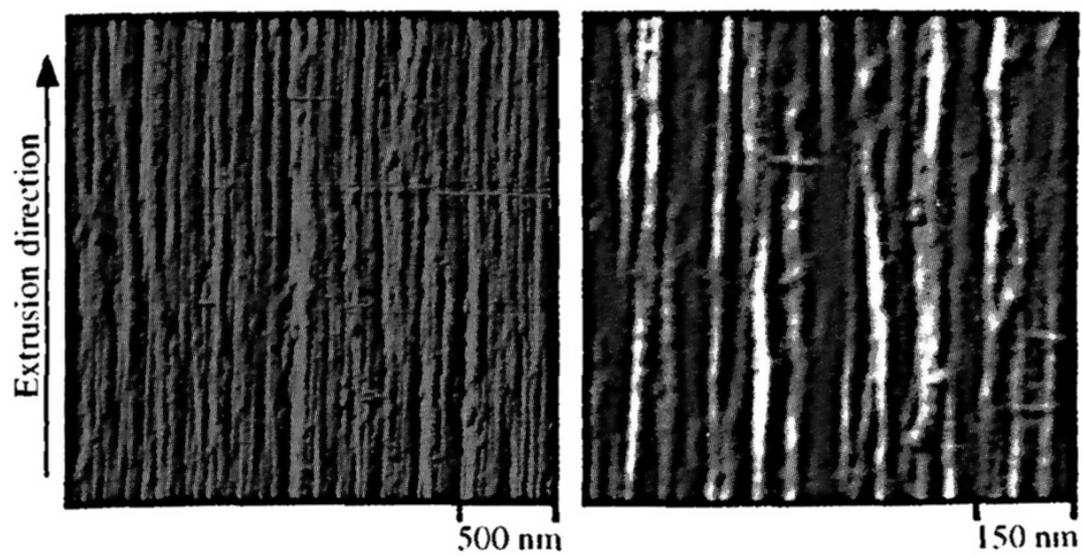

## Figure 10

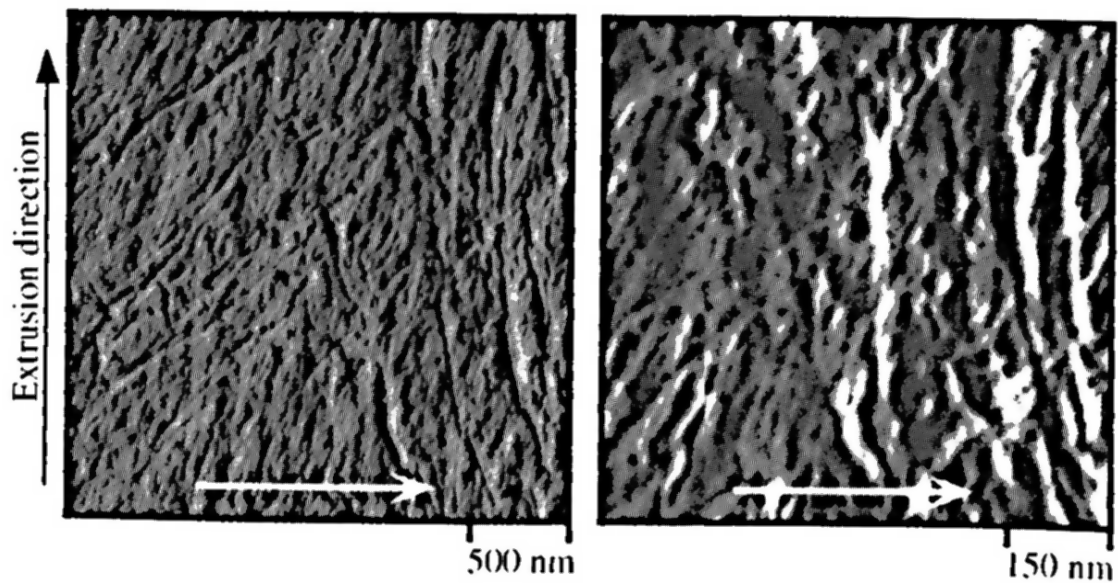

## Figure 11

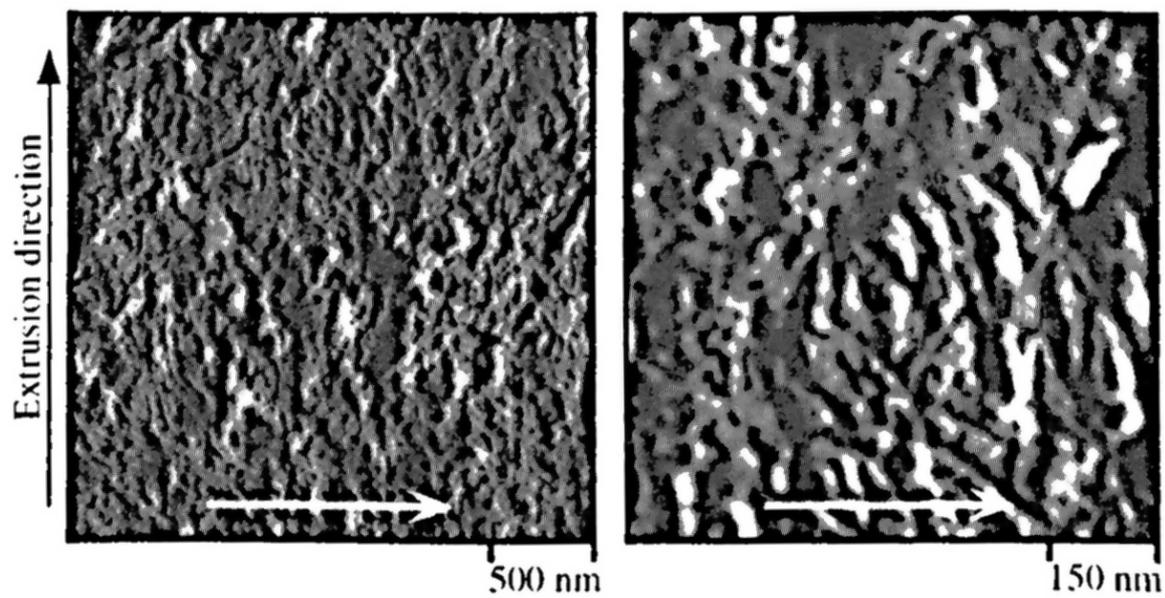

## Figure 12

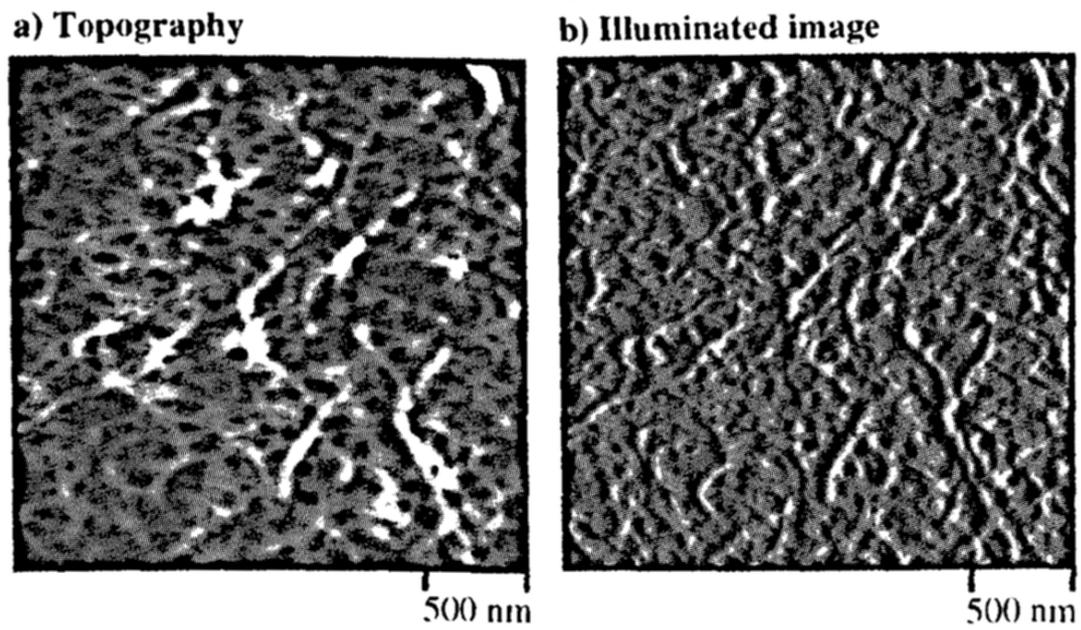

## Figure 13

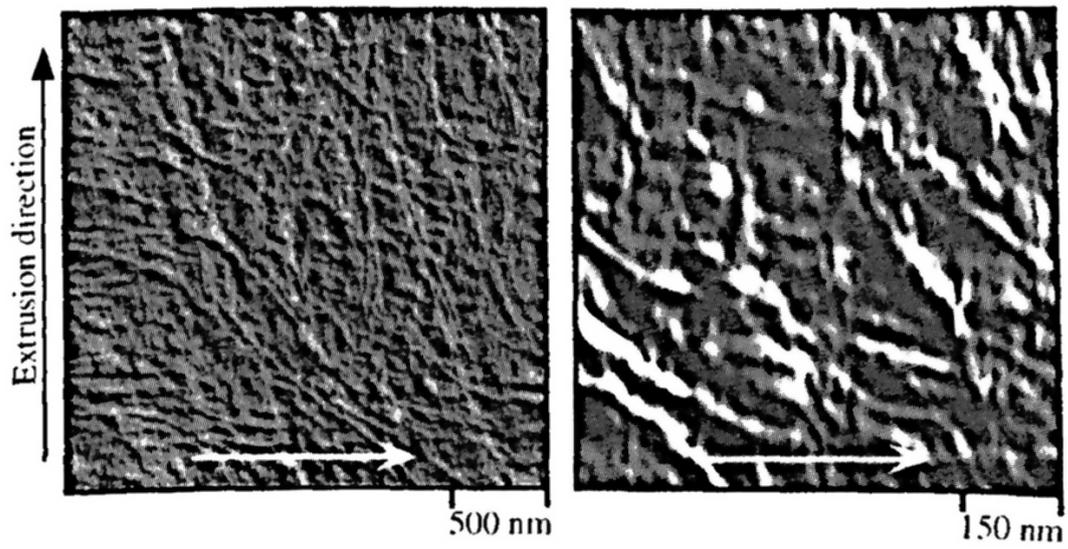

# Figure 14

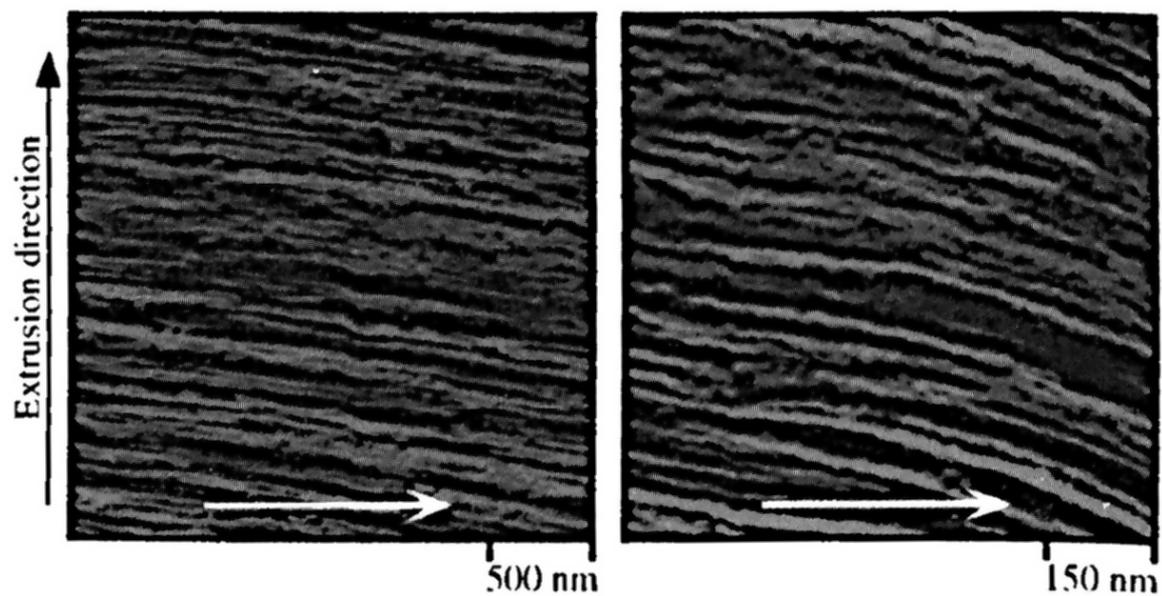

# Figure 15

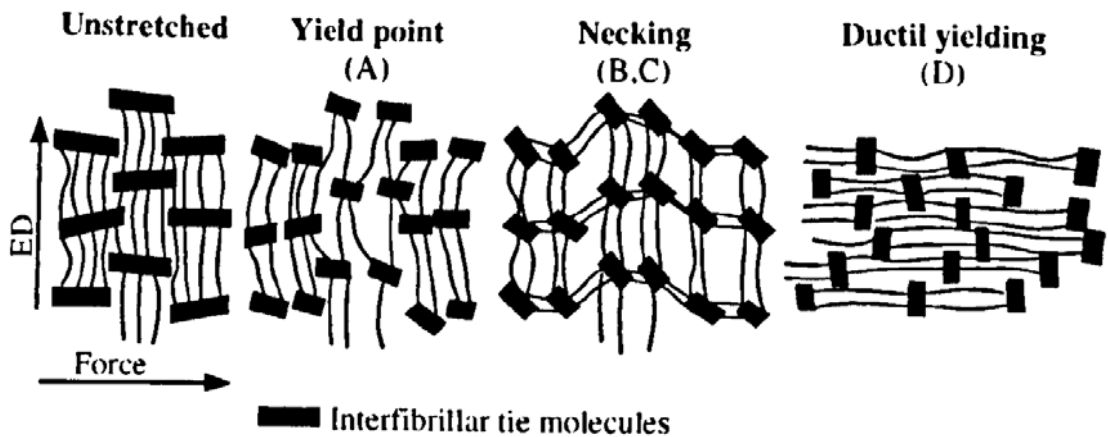

# Figure 16

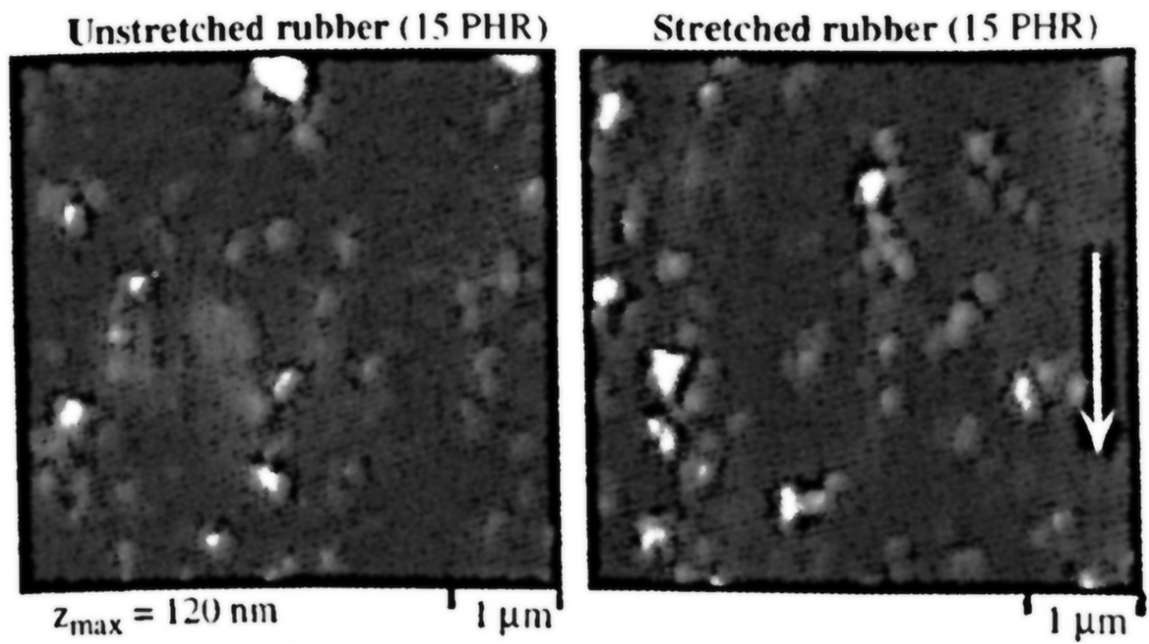

Appeared in Ratner, B. D. & Tsukruk, V. V. (Eds.) "Scanning Probe Microscopy of Polymers" as Chapter 6, American Chemical Society and Oxford University Press, 1998, 694, 110-128

## Figure 17

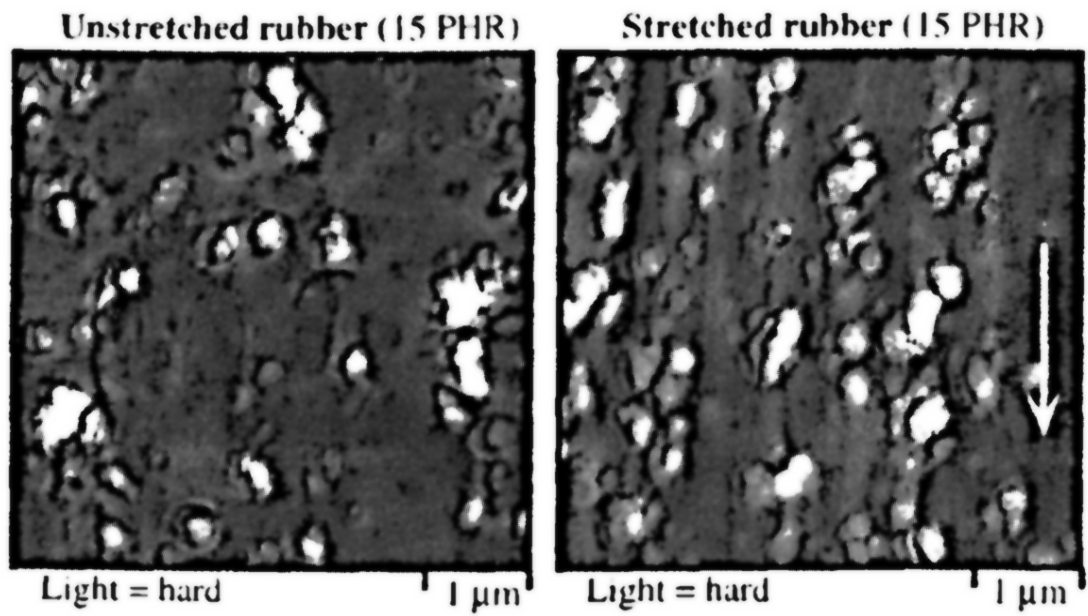

Figure 18

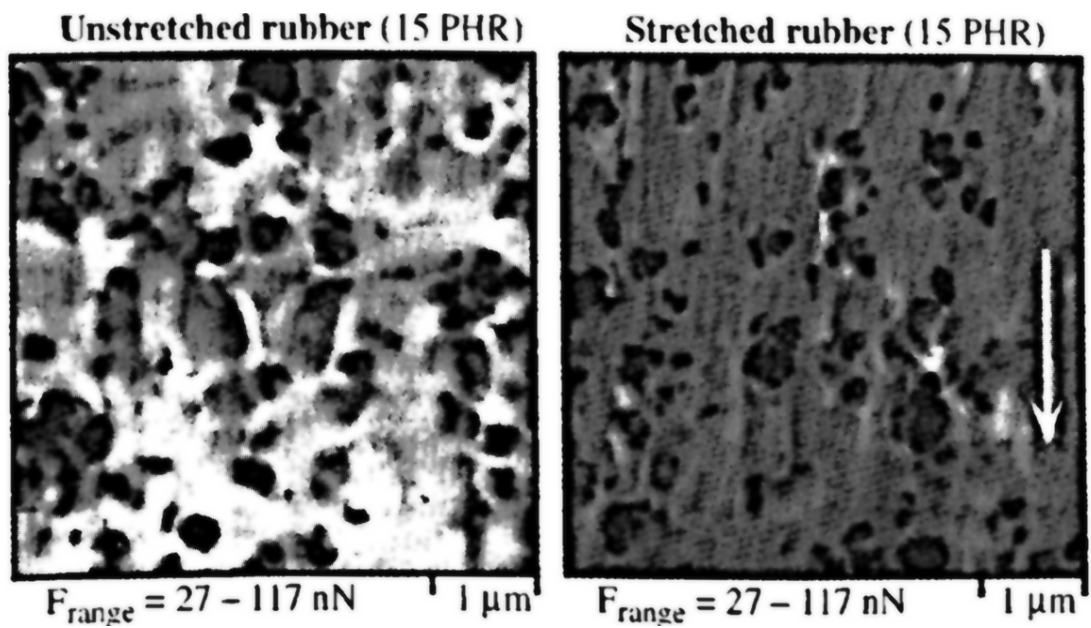

Figure 19

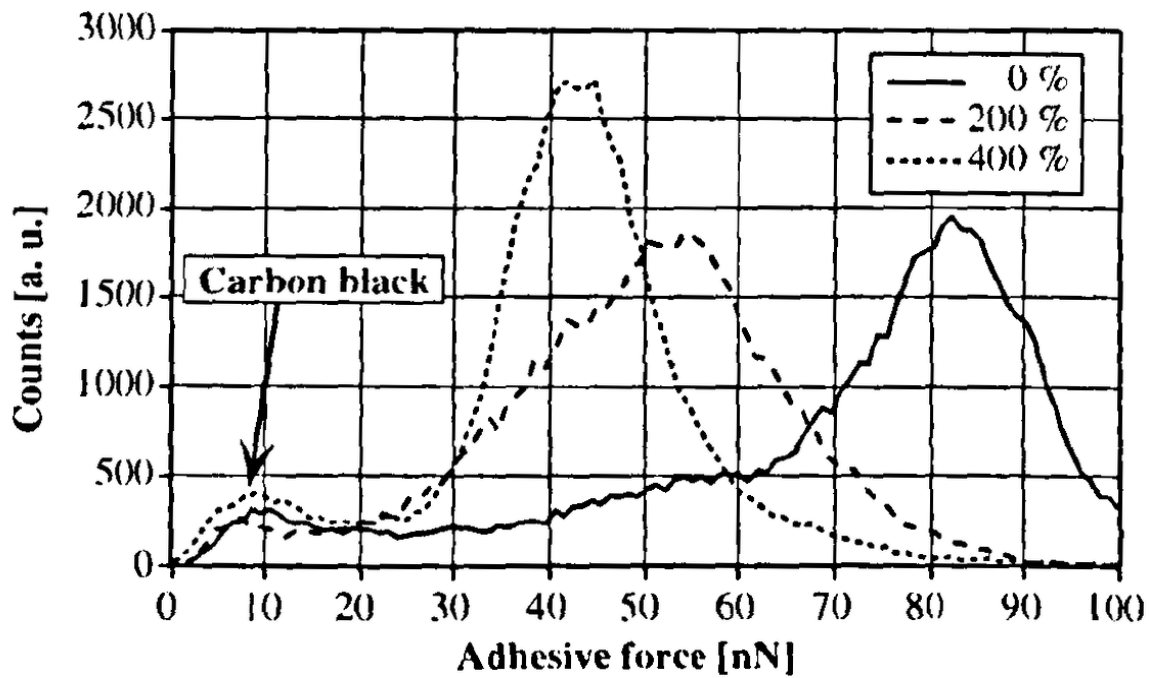